\documentclass[12pt]{article}
\usepackage{epsf}
\usepackage{amsmath}
\usepackage{amssymb}
\usepackage{graphicx}

\title{
\begin{flushright}
\normalsize\rm TPJU-11/2006
\end{flushright}
\vskip 10 pt
Large $N$ behavior of two dimensional supersymmetric Yang-Mills quantum mechanics}
\author{Maciej Trzetrzelewski
\footnote{trzetrzelewski@th.if.uj.edu.pl }   \\
 \emph{M. Smoluchowski Institute of Physics, Jagiellonian
University} \\
\emph{Reymonta 4, 30-059 Krak\'ow, Poland}
}

\begin{document}
\date{}

\maketitle

\abstract{We analyze the $N \to \infty $ limit of
supersymmetric Yang-Mills quantum mechanics (SYMQM) in two
spacetime dimensions. To do so we introduce a particular class of
$SU(N)$ invariant polynomials and give the solutions of 2D SYMQM
in terms of them. We conclude that  in this limit the system is
not fully described by the single trace operators
$Tr({a^{\dagger}}^n)$ since there are other, bilinear operators
$Tr^n(a^{\dagger}a^{\dagger})$ that play a crucial role when the
hamiltonian is free.}
\vspace{0.5cm}


\section{Introduction}

Since the B.F.S.S matrix conjecture [1] relating M-theory with
supersymmetric Yang-Mills quantum mechanics in  D=9+1 dimensions,
there has been a lot of interest in solving the above quantum
mechanical systems and their lower dimensional relatives. They are
governed by the hamiltonian [2]

\begin{equation}
H= \frac{1}{2}\pi^i_a \pi^i_a + \frac{1}{4}g^2 (f_{abc} x^i_b x^j_c)^2  +H_F,
\end{equation}

\noindent where $a,b,c=1, \ldots, N^2-1$ ale color indices  of
$SU(N)$ group, $i,j=1,\ldots,D-1$ are spatial indices and  $H_F$
is the fermionic term the details of which will not be explored
here. The zero energy states of (1) have been widely  investigated
in number of papers [3] which mainly focus on the existence and
asymptotic form of ground states.
 In D=1+1 case the exact supermultiplet structure
for $SU(2)$ group were known for a long time due to Claudson
and Halpern [2] . Later on they were extended for arbitrary
$SU(N)$  in purely bosonic sector [4]. It is not surprising that
D=1+1 systems are solvable since the quartic potential in (1)
vanishes in two dimensional systems. The hamiltonian is not free
however due to the Gauss law. In other words all physical states obey the
constraint $G_a\mid s \rangle=0$ where $G_a$ are the $SU(N)$
generators. There are no exact solutions in higher dimensions but
a huge effort has been made to study them numerically. The D=3+1
system in bosonic sector has its origin in zero volume pure
Yang-Mills theory [5]. Its up to date study for $SU(2)$ gauge
group can be found in [6] where  nonperturbative values of the
spectrum, the Witten index, the supermultiplet structure and the
wave functions are discussed.  In D=9+1 even for $SU(2)$
case there are no analogous calculations ( i.e. numerical ones )
due to the high complexity of the system.

In this paper we deal with the large $N$ behavior  of the D=1+1
systems hoping the the analogous approach will prove useful in
higher dimensions. The approach presented here differs  from the
ones existing in the literature e.g. [7]. We will prove that in $N
\to \infty$ limit the solutions admit very simple form hence it is
possible that
 the same thing takes place in higher
dimensions. We also discuss the recent work by Veneziano and
Wosiek  [8] where planar quantum mechanics is studied as well
although in terms of the different model.

The structure of the paper is the following. In  section 2 we
review the Claudson-Halpern-Samuel solution. In sections 3 and 4,  using the algebraic approach,
we obtain a new set of solutions for which the $N$ dependence is
evident so that the large $N$ limit is manageable. In section  5 we
discuss the vacuum structure of the models. Finally in section 6
we proceed with the $N \to \infty$ limit for both free and
harmonic oscillator hamiltonians. We also point out the relevance
of our solutions in that limit.

\section{Claudson-Halpern-Samuel solutions}

The two dimensional system is described by real bosonic coordinates
$x_a$ and conjugate momenta $\pi_a$, $[x_a,\pi_b]=i\delta_{ab}$
and fermionic variables  $f_a$, $f^{\dagger}_a$,
$\{f_a,f^{\dagger}_b\}=\delta_{ab}$. The hamiltonian and the
supercharges are then [2]

\[
H=\frac{1}{2}\pi_a \pi_a+gx_aG_a, \ \ \ \ Q=f_a \pi_a,
\]
where $G_a=f_{abc}(x_b \pi_c-if^{\dagger}_b f_c)$ is the $SU(N)$
generator, $f_{abc}$ are $SU(N)$ structure tensors. The physical
states $\mid s \rangle$ are those obeying the gauss law $G_a\mid s
\rangle=0$ i.e. they are $SU(N)$ singlets. In this $SU(N)$
invariant subspace the hamiltonian is supersymmetric and free i.e
$2H=\{Q,Q^{\dagger}\}=\pi_a \pi_a\equiv \pi^2$.

The first solutions and the hole supermultiplet structure were obtained
by Claudson and Halpern [2] for $SU(2)$ group. In bosonic sector
they are

\[
\mid k \rangle= \frac{sin(k r)}{k r}\mid v \rangle,   \ \ \ \
\pi^2\mid k \rangle=k^2\mid k \rangle, \ \ \ \ r=\sqrt{x_ax_a},
\]
where $\mid v \rangle$ is the vacuum state  $Q^{\dagger}\mid v
\rangle=Q\mid v \rangle=0$ \footnote{ In this paper we denote the supersymmetric vacuum as $\mid v
\rangle $ while the Fock vacuum as $\mid 0
\rangle $. }. The generalization of these solutions
for arbitrary gauge group $U(N)$ is due to Samuel [4].  The idea is
to work with $U(N)$ invariants $\lambda_i$, $i=1,\ldots,N $ i.e.

\[
X=UDU^{-1}, \ \ \ \ D=diag(\lambda_1,\ldots,\lambda_N),
\]
and search for the solutions of the form
$f(\lambda_1,\ldots,\lambda_N)$. The eigenequation in this
coordinates is then

\[
\pi^2f=-\frac{1}{M^2}\frac{\partial}{\partial \lambda_i}M^2
\frac{\partial}{\partial \lambda_i}f=k^2f, \ \ \ \
M=\prod_{i<j}(\lambda_i-\lambda_j),
\]
and the solution is
\[
f(\lambda_1,\ldots,\lambda_N)=\frac{1}{M}exp(i\sum_{j} k_j \lambda_j).
\]
These solutions behave badly as $\lambda_i$ approaches
$\lambda_j$, $i \ne j$ so one has to consider their superposition
( antisymmetrization )
\[
F(\lambda_1,\ldots,\lambda_N)=\frac{1}{N!}\sum_{\sigma \in  S_N
}(-1)^{sgn(\sigma)}f(\lambda_{\sigma(1)},\ldots,\lambda_{\sigma(N)}),
\ \ \ \ \pi^2 F=k^2F.
\]
These solutions are now regular. Note that Samuel's solutions can
be generalized making the following anzatz

\[
f(\lambda_1,\ldots,\lambda_N)=\frac{1}{M}g(\lambda_1,\ldots,\lambda_N).
\]
The differential equation $ \pi^2f=k^2f $ gives $\nabla^2g+k^2g=0$
which is the Helmholtz differential equation in $N$ dimensions. It
is now evident that Samuel's solutions are special ones for which
$g=exp(i\sum_j k_j \lambda_j)$, $k^2=\sum_j k_j^2$, i.e. they
correspond to the plane-wave solutions of the Helmholtz equation. It
seems that Samuel's solutions provide a natural basis of all the
solutions since the superposition $\int h(k_1,\ldots,k_N)exp(i\sum_j
k_j \lambda_j)$  obeys Helmholtz equation as well provided
$k^2=\sum_j k_j^2$.  \footnote{ If there are boundary conditions eg.
$grad_{\vec{n}}g=0$ where $\vec{n}$ is a normal vector to the $N-1$
dimensional, closed surface, then the situation is more subtle. It
is not clear what, if any, boundary conditions we should take in
this case therefore we do not discuss it.} However we will indicate
in section 4 that there exist solutions when this does not happen.

Note that in this approach the $N$ dependence is not explicit
therefore we change the variables from $\lambda_k$ to $(X^k)
\equiv Tr(X^k)$ in the following way.  The  general solution
$\frac{1}{M}g$ has to be antisymmetrized due to the $1/M$ factor.
This makes the solution completely symmetric and therefore it can
be expanded in terms of symmetric polynomials
$\sum_i\lambda_i=Tr(X)=0$ ( for $SU(N)$ ),
$\sum_i\lambda_i^2=Tr(X^2) \equiv (X^2)$,
$\sum_i\lambda_i^3=Tr(X^3)\equiv(X^3)$ etc., so that general
solutions can be written as

\begin{equation}
\mid s \rangle =\sum c_{i_2\ldots i_N}(X^2)^{i_2}\ldots (X^N)^{i_N} \mid v \rangle.
\end{equation}
We will now attempt to reconstruct these solutions  ( i.e.
determine the $c_{i_2\ldots i_N}$'s )  by  algebraic methods . It
will soon appear clear that  the algebraic approach gives the
possibility to study the $N \to \infty$ limit.

\section{An algebraic approach}

In this section we use the following conventions

\[
T_iT_j=\frac{1}{N}\delta_{ij}\mathbf{1}+\frac{1}{\sqrt{2}}(d_{ijk}+if_{ijk})T_k,
\]

\noindent where $T_i$'s are $su(N)$   generators in the
fundamental representation  and $f_{ijk}$/$d_{ijk}$  are
complectly antisymmetric/symmetric structure tensors.  Moreover, we do not
specify here the representation of momentum $\pi_a$ and coordinate
$x_a$ operators. All we need is their commutation relation
$[x_a,\pi_b]=i\delta_{ab}$.

Note that the  conventions that we use differ from the
common ones appearing in the literature, namely
\[
T_iT_j=\frac{2}{N}\delta_{ij}\mathbf{1}+(d_{ijk}+if_{ijk})T_k.
\]
There is a technical reason for doing so.  First, if we define
$(X^2)=x_ax_bTr(T_aT_b)$ then we have $(X^2)=x_ax_a$ instead of
$(X^2)=2x_a x_a$ with an awkward factor of 2. Moreover the
standard identity for $SU(N)$ generators is now
\[
[T_a]_{ij}[T_a]_{kl}=\delta_{il}\delta_{jk}-\frac{1}{N}\delta_{ij}\delta_{kl},
\]
instead of
\[
[T_a]_{ij}[T_a]_{kl}=2\delta_{il}\delta_{jk}-\frac{2}{N}\delta_{ij}\delta_{kl}.
\]
We shall also use the notation
\[
(T_aT_b \ldots)\equiv Tr(T_aT_b \ldots), \ \ \ \ (AB\ldots)=A_aB_b\ldots(T_aT_b\ldots).
\]

Let us consider the most general form of the $SU(N)$ invariant
eigenstates in purely bosonic sector

\begin{equation}
\mid s \rangle = \sum T_{acb\ldots}x_ax_bx_c\ldots\mid v \rangle,
\end{equation}
where $T_{abc\ldots }$  is some $SU(N)$ invariant tensor made out
of $SU(N)$ tensors $f_{ijk}$, $d_{ijk}$ and $\delta_{ij}$.
The contractions between these structure tensors may be arbitrary
so the best way to imagine them is to work  in Cvitanovi$\check{c}$
notation [9] (figure 1)

\begin{figure}[h]
\centering \leavevmode
\includegraphics[width=0.6\textwidth]{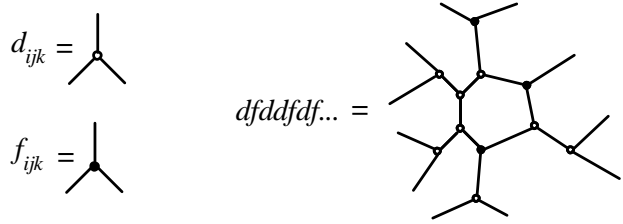}
\caption{$d_{ijk}$, $f_{ijk}$ diagrams and a typical tensor
diagram.}
\end{figure}
\noindent Each tensor is now represented by a diagram which in
general has many loops. One can prove that these loops are not
present [10] i.e. they are linear combination  of trees. Tree
diagram may be expressed in terms of products of trace tensors
$Tr(T_a T_b\ldots)$ so if one introduces the following matrix
$X=x_a T_a$ and $(X^k)\equiv Tr(X^k)$ then using the
Cayley-Hamilton theorem for matrices $X$, Eqn. (3) becomes
\begin{equation}
\mid s \rangle = \sum c_{i_2\ldots i_N}(X^2)^{i_2}\ldots (X^N)^{i_N} \mid v \rangle,
\end{equation}
which coincides with (2) as it should.

If $\mid s \rangle$ is an eigenstate of the hamiltonian
$H=\frac{1}{2}\pi^2$ then the coefficients $c_{i_2\ldots i_N}$ can
be determined with use of the following commutation relations (
see Appendix A )

\begin{equation}
[\pi^2,(X^2)^n]=-4in(X^2)^{n-1}(X\pi)-2n(2n+1+N^2-4)(X^2)^{n-1},
\end{equation}

\begin{equation}
[\pi^2,(X^k)]=-2ik(X^{k-1}\pi)-k(2N-\frac{k-1}{N})
(X^{k-2})- \epsilon k\sum_{j=2}^{k-4} (X^{k-2-j})(X^j),
\end{equation}
where $\epsilon=0$ for $k=3,4,5$ and $\epsilon=1$ for $k>5$.
Using (5) and (6) one can prove ( see Appendix B )  that there
exist polynomials $P_k \equiv P_k((X^2),(X^3),\ldots,(X^k))$ of
order k in variables $x_a$  such that \footnote{The  order of the polynomial
is understood here as the maximal number of times one has to
differentiate the polynomial to make it vanish. For example
polynomial $x_1x_2x_3$ is of degree 3 although all the variables
have degree 1 independently. }

\begin{equation}
[\pi^2,P_k]=-2i\partial_aP_{k} \ \pi_a, \ \ \ \  P_k
\xrightarrow[N \longrightarrow \infty]{}  (X^k), \ \ \ \
x_a\partial_aP_{k}=kP_k, \ \ \ \  \Delta P_k=0, \ \ \ \ k=3,4,\ldots  \ \ \ .
\end{equation}
The first few $P_k$'s are

\[
P_3=(X^3), \ \ \ \ P_4=(X^4)-\frac{4N-\frac{9}{2N}}{N^2+1}(X^2)^2.
\]
Using (5) and (7) we have

\[
\pi^2 (X^2)^n \mid v \rangle=-2n(2n+1+N^2-4) (X^2)^{n-1} \mid v \rangle,
\]

\[
\pi^2 P_k (X^2)^n \mid v \rangle=-2n(2n+1+N^2-4+2k)P_k (X^2)^{n-1} \mid v \rangle,
\]
therefore we obtain a class of solutions ( up to the normalization factor )
\begin{equation}
\mid p \rangle=\frac{1}{pr}\sin_{N^2-4}(pr)\mid v \rangle, \ \ \ \ \pi^2\mid p \rangle=p^2\mid p \rangle,
\end{equation}

\begin{equation}
\mid p \rangle_k=\frac{P_k}{pr}\sin_{N^2-4+2k}(pr)\mid v \rangle, \ \ \ \ \pi^2\mid p \rangle_k=p^2\mid p \rangle_k,
\end{equation}
where
\[
r\equiv\sqrt{x_ax_a}=\sqrt{(X^2)},
\]
and
\[
\sin_t(x)=\sum_{k=0}^{\infty} \frac{x^{2k+1}}{1 \cdot 2 (3+t)4(5+t)\ldots 2k(2k+1+t)}.
\]

\section{Recurrence relations and initial conditions}

Solutions (8,9) are not general ones since we see from equation (4)
that the general solution may have arbitrary powers of $(X^k)$
operators which is not the case in (8,9). This problem is not
present for $SU(2)$ group since then the differential eigenequation
is with respect to only one variable and the Claudson-Halpern
solutions are already the general ones. Indeed we confirm that with
algebraic approach since for $SU(2)$ the general form of a solution
is

\begin{equation}
\mid p \rangle=\sum_{n}a_n (X^2)^n \mid v \rangle, \ \ \ \ \pi^2
\mid p \rangle=p^2\mid p \rangle,
\end{equation}
Applying (5) to (10) we obtain the following recurrence for $a_n$
\[
p^2 a_{n-1}=-a_n 2n(2n+1) \Rightarrow a_n=(-1)^n\frac{a_0 p^{2n}}{(2n+1)!}.
\]
Introducing $r^2=(X^2)$ we obtain the Claudson-Halpern solution
\[
\mid p \rangle=a_0 \frac{1}{pr}sin(pr)\mid v \rangle,
\]
with the normalization factor $a_0$.

Let us now apply the same method for the $SU(3)$ group. The general
solution is

\begin{equation}
\mid p \rangle=\sum_{n}a_{nm} (X^2)^n (X^3)^m \mid v \rangle, \ \
\ \ \pi^2 \mid p \rangle=p^2\mid p \rangle.
\end{equation}
Applying (5,6) to (11) we get two recurrence relations for $a_{nm}$

\footnote{ To derive (13) we used the identity
\[
\pi^2(X^2)^n(X^3)^m \mid v
\rangle=-2n(2n+1+6m+5)(X^2)^{n-1}(X^3)^m \mid v \rangle
-\frac{2}{3}m(m-1) (X^2)^{n+2}(X^3)^{m-2}\mid v \rangle,
\]
which follows from (29) and $(XXT_a) (XXT_a)=(X^4)-\frac{1}{3}(X^2)^2=\frac{1}{6}(X^2)^2$.}

\begin{equation}
p^2a_{n-1, m}=-2n(2n+1+6m+5)a_{n, m}, \ \ \ \ m=0,1,
\end{equation}
\begin{equation}
p^2a_{n-1, m}=-2n(2n+1+6m+5)a_{n, m}-\frac{2}{3}(m+1)(m+2)a_{n-3, m+2} \ \ \ \ m>1.
\end{equation}
Note that (12) already gives us two independent, diagonal in $m$, solutions namely

\begin{equation}
\mid p_2 \rangle =\frac{a_{00}}{p_2 r}sin_{5}(p_2 r)\mid v
\rangle, \ \ \ \ \pi^2\mid p_2 \rangle={p_2}^2\mid p_2 \rangle,
\end{equation}
and

\begin{equation}
\mid p_3 \rangle =\frac{(X^3)a_{01}}{p_3 r}sin_{11}(p_3 r)\mid v
\rangle, \ \ \ \ \pi^2\mid p_3 \rangle={p_3}^2\mid p_3 \rangle.
\end{equation}
They are exactly the solutions (8,9)  for $SU(3)$. In terms of 2D
lattice where each point $(n,m)$ represents a vector
$(X^2)^n(X^3)^m \mid v \rangle $ it means that the solutions $\mid
p_2 \rangle$, $\mid p_3 \rangle$ exist on two rows $(n,0)$ and
$(n,1)$. The recurrence (13) is represented in figure 2.

\begin{figure}[h]
\centering \leavevmode
\includegraphics[width=0.3\textwidth]{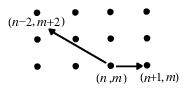}
\caption{The structure of recurrence (13).}
\end{figure}
 As already indicated the solutions (8,9) are not general ones. It follows that if one
fixes e.g. the set $A_2=\{ a_{n 2}, n>0\} $ or the set $A_3=\{
a_{n 3}, n>0\} $ then (13) determines all $a_{n, 2m}$, $m>0$ or
$a_{n, 2m+1}$, $m>0$ respectively. We will denote the solutions
corresponding to these coefficients as $ \mid p_{A_2} \rangle$, $
\mid p_{A_3} \rangle$ (figure 3 ). Moreover we see that the
solutions (8,9) correspond to trivial initial conditions i.e. to $a_{n 2}=a_{n 3}=0$ far all $n$. The general
solution is now
$$\mid p_{1} \rangle+\mid p_{2} \rangle+\mid p_{A} \rangle+\mid p_{B} \rangle,$$
with the energy
$$p_1^2+ p_2^2+p_{A_2}^2+p_{A_3}^2.$$

\begin{figure}[h]
\centering \leavevmode
\includegraphics[width=0.4\textwidth]{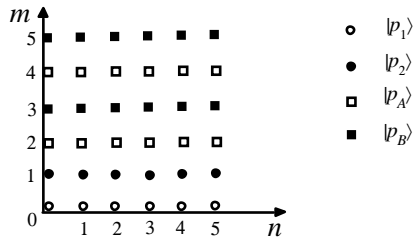}
\caption{Four independent solutions represented on lattice.}
\end{figure}

The case of $SU(N)$ is analogous. There are solutions (8,9)
corresponding  to trivial initial condition \footnote{There may
exist solutions different from (8,9) that also correspond to
trivial initial condition. We do not discuss them here.   } and
other solutions that require to set the infinite amount of initial
coefficients for the recurrence analogous to (13).

Let us note that Samuel's solutions require to set the infinite
amount of coefficients. Only then it is possible to obtain the
plane wave from the  corresponding  recurrence. It follows that
there are plenty of solutions, e.g. Eqns (8,9), that cannot be
obtained from the superposition of Samuel's solutions.

\section{Vacuum states}
Here we construct the supersymmetric vacuum state  $\mid v
\rangle$: $Q^{\dagger}\mid v \rangle=Q\mid v \rangle=0$ in several
fermion sectors. For $n_F=0$ sector the general form of the vacuum
state is
\[
\mid v \rangle=\sum c_{i_2 i_3 \ldots i_N}
({a^{\dagger}}^2)^{i_2}({a^{\dagger}}^3)^{i_3}\ldots
({a^{\dagger}}^N)^{i_N}\mid 0 \rangle, \ \ \ \  Q^{\dagger}\mid v
\rangle=0,
\]
where $\mid 0 \rangle$ is the Fock vacuum  and
$({a^{\dagger}}^k)=Tr({a^{\dagger}}^k)$,
$a^{\dagger}=a^{\dagger}_bT_b$. The supersymmetric generator
$Q^{\dagger}$ in terms of creation and annihilation operators is
$i\sqrt{2}Q^{\dagger}=(af^{\dagger})-(a^{\dagger}f^{\dagger})$
therefore the condition for vacuum becomes
\begin{equation}
(af^{\dagger})\mid v \rangle=(a^{\dagger}f^{\dagger})\mid v \rangle.
\end{equation}
The last equation can be satisfied only if
\[
\mid v \rangle=\sum_{k=0}^{\infty} c_{k} ({a^{\dagger}}^2)^k\mid 0 \rangle.
\]
This can be proved in the following way.  Suppose that the vacuum
$\mid v \rangle$ contains traces $(a^k), k>2$.  Since
$(af^{\dagger})({a^{\dagger}}^k)\mid 0
\rangle=k(f^{\dagger}{a^{\dagger}}^{k-1})\mid 0 \rangle$, the left
hand side of (16) will certainly contain traces
$(f^{\dagger}{a^{\dagger}}^{k-1})$ which are not present on the
right hand side of (16). Therefore there are no traces $({a^{\dagger}}^k),
k>2$.  For $k=2$ the situation is different because the operator
$({a^{\dagger}}f^{\dagger})$ appears in the right hand side of
(16). Now it is straightforward to prove that
$c_{k}=\frac{1}{(2k)!!}$
therefore in $n_F=0$ sector there is only one vacuum given by
\[
 \mid v \rangle=\sum_{k=0}^{\infty} \frac{1}{(2k)!!} ({a^\dagger}^2)^k\mid 0 \rangle.
\]
The above formula generalizes the $SU(2)$ case [11] to arbitrary $SU(N)$.
Note however that this state is badly normalized \footnote{ We use the relation analogous to (5) i.e.
\[
[(a^2),({a^{\dagger}}^2)^n]=4n({a^{\dagger}}^2)^{n-1}(a^{\dagger}a)+2n(2n-2+N^2-1)({a^{\dagger}}^2)^{n-1}.
\]
}

\[
 \langle v \mid v \rangle=\sum_{k=1}^{\infty} (1+\frac{N^2-3}{2})(1+\frac{N^2-3}{4})\ldots(1+\frac{N^2-3}{2k}).
\]
This is as it should be since the theory is free.  It is however
the source of some inconsistencies that we will investigate later.

 Now, we can construct the remaining vacuum states in
$n_F>0$ sectors. They are made out of fermion operators
$({f^{\dagger}}^{k})$ acting on the state $\mid v \rangle$. Most
of operators $({f^{\dagger}}^{k})$ vanish. The list of nontrivial
and independent ones is ( see Appendix C )

\begin{equation}
({f^{\dagger}}^{3}), \ \ \ ({f^{\dagger}}^{5}), \ldots,
({f^{\dagger}}^{2N-1}) \ \ \ \ \ \hbox{for $SU(N)$}.
\end{equation}
Such operators are nonvanishing, nilpotent and they commute with
$p_a$  therefore defining
\begin{equation}
\mid \epsilon_3, \ldots,\epsilon_{2N-1}  \rangle:=({f^{\dagger}}^{3})^{\epsilon_3}
({f^{\dagger}}^{5})^{\epsilon_5}\ldots({f^{\dagger}}^{2N-1})^{\epsilon_{2N-1}}\mid
v \rangle, \ \ \ \ \epsilon_i=0,1,
\end{equation}
we have
\[
Q^{\dagger}\mid \epsilon_3, \ldots,\epsilon_{2N-1}   \rangle=Q\mid
\epsilon_3, \ldots,\epsilon_{2N-1}  \rangle=0,
\]
hence $\mid \epsilon_3, \ldots,\epsilon_{2N-1}  \rangle$ are vacuum states.

We immediately see that the number of such states, in each fermion
sector, is given by the generating function

\begin{equation}
(1+t^3)(1+t^5)\ldots(1+t^{2N-1})=\sum_{i=0}^{N^2-1} b_i t^i,
\end{equation}
i.e. the number of vacuum states  (18) in $n_F$ sector is given by
$b_{n_F}$. We also recognize the polynomial (19) as the generating
function for Betti numbers ( i.e Pioncar\'e polynomial ) of $SU(N)$
group manifold [12] while the vacuum states (18) correspond to
nontrivial Lie algebra cocycles of $SU(N)$ [14]. A possible relation
of this model to Witten's quantum mechanics [13] on $SU(N)$ manifold
will be discussed elsewhere. Instead let  us return to the
non-normalization of the vacuum state. It is a source of some
inconsistencies that we will now discuss \footnote{I thank G.
Veneziano for discussions on this subject }.

 In section 3 we derived the exact formula
for the solutions with trivial initial condition Eqns.
( 8,9 ). Remarkably they do not vanish if we put $p=0$. In fact we
obtain

\begin{equation}
\mid p=0 \rangle_k=P_k \mid v \rangle, \ \ \ \ \pi^2\mid p=0 \rangle_k=0,
\end{equation}
therefore we obtain a countable set of zero  energy solutions in
sector with no fermions. In other fermion sectors we have the same
situations, i.e. the states

\[
\mid \psi \rangle_k=P_k\mid \epsilon_3, \ldots,\epsilon_{2N-1}   \rangle , \ \ \ \ k>2, \ \ \ \ \epsilon_i=0,1,
\]
are zero energy states as well. Note however  that while $ \pi^2\mid
\psi \rangle_k=0$ we also have $Q^{\dagger}\mid \psi \rangle_k \ne
0$ !  It means that the basic theorem in supersymmetry, namely

\begin{equation}
Q^{\dagger}\mid v \rangle =0 \Longleftrightarrow H\mid v \rangle=0,
\end{equation}
does not hold for $H=\frac{1}{2}\pi^2$ in the $SU(N)$ invariant
sector\footnote{ For  example $k=3$ gives $P_3=(X^3)$ then
\[
\pi^2(X^3)\mid v \rangle=-6i(X^2\pi)\mid v \rangle=0,
\]
but
\[
Q^{\dagger}(X^3)\mid v \rangle=-3i(X^2f^{\dagger})\mid v \rangle \ne 0.
\]
}. To be more precise the  implication $\Rightarrow$ is formally
correct but the converse is not!  This inconsistency is due to the
bad normalization of the vacuum state. This state to be
mathematically correct does not exist in the Hilbert space therefore
the proof of the theorem (21), which assumes that $\langle v \mid Q
Q^{\dagger} \mid v \rangle <\infty$, does  not hold anymore. The
remedy at this point is to compactify $x_a$ so that all the states
are normalizable. It turns out that then  the model discussed admits the topological interpretation. This issues
will be discussed elsewhere.

 To conclude this section we
see that for $SU(N>2)$ there are  two classes of zero energy
states $\mid \psi \rangle$ and $\mid \psi \rangle_k$, $H\mid \psi
\rangle=H\mid \psi \rangle_k=0$ but only the first class
corresponds to supersymmetric vacuum i.e $Q^{\dagger}\mid \psi
\rangle=0$ while $Q^{\dagger}\mid \psi \rangle_k \ne 0$.

\section{The large N limit}
Taking the $N \to \infty$ limit in Eqns. (8,9) gives
\begin{equation}
\mid p \rangle_k  \xrightarrow[N \longrightarrow \infty]{}  (X^k) \mid v \rangle,
\end{equation}
where the above limit is understood in the sense of norms i.e.
$\lim_{N \to \infty} \left\| \mid p \rangle_k \right\|=\lim_{N \to
\infty} \left\| (X^k) \mid v \rangle \right\|$. Thus the $N=\infty$
case is obtained by the operators  $(X^k)$ acting on supersymmetric
vacuum.

Two comments are now in order. Note that these states also have
zero energy since in large $N$ limit (20)  coincide with (22).
Therefore the states from the continuous spectrum collapse into
the zero energy states in the $N \longrightarrow \infty$ limit.
It is as if the low energy behavior of the finite $N$ system was given by
its large $N$ limit.  The lack of $p$ dependence on the right hand
side of Eqn. (22) is due to the fact that we are considering  the
$N \to \infty$ limit of particular solutions. It does not mean
that in this limit there are no solutions with $p$ dependence.
Such solutions do exist but they correspond to some nontrivial initial
conditions ( see section 4). Moreover the $N=\infty$ case describes no
more a quantum mechanical system but rather quantum field theory
since the number of degrees of freedom is now infinite. It is now
tempting to interpret the infinite number of zero energy states
(22) as massless states in some emerging quantum field theory, at
least we see that in this limit the hamiltonian $H=\frac{1}{2}\pi^2$ provides
such  possibility.

Let us now rewrite  conclusion (22) in terms of creation operators
acting on the Fock vacuum $\mid 0 \rangle$. It requires little work to
prove ( see Appendix C ) that in the large $N$ limit vectors (22) are
given by the linear combinations of states

\begin{equation}
\mid n,m \rangle = ({a^{\dagger}}^2)^n ({a^{\dagger}}^m) \mid 0 \rangle.
\end{equation}
This result is important since it means that in the large $N$ limit the
passage from  coordinate operators to creation operators  is not realized
simply by the substitution $X \rightarrow a^{\dagger}$,  $\mid v
\rangle \rightarrow \mid 0 \rangle$. The difference lies in the
structure of the vacuum state in Fock space. In recent work on
planar quantum mechanics [8] Veneziano and Wosiek argued that the
most important states in Fock space are those given by single
trace i.e.

\begin{equation}
\mid n \rangle = ({a^{\dagger}}^n)\mid 0 \rangle,
\end{equation}
therefore vectors (24) differ from (23)  by the absence of
$({a^{\dagger}}^2)^n$ operators which are exactly due to the
vacuum structure in Fock space. If we want to take
that into consideration we should rather work with the basis
\begin{equation}
\mid n \rangle = ({a^{\dagger}}^n)\mid v \rangle,
\end{equation}
then with (24).

At this stage one can also ask weather the  basis (23) is
characteristic of the $N \to \infty$ basis for  the free hamiltonian
or is it a good basis even if bound states occur. We verify this
question explicitly on the example in supersymmetric harmonic
oscillator. The hamiltonian and the supersymmetric charge are
respectively
\[
H=\frac{1}{2}\{Q,Q^{\dagger}\}=a^{\dagger}_ba_b+f^{\dagger}_b f_b=\frac{1}{2}(\pi_b\pi_b+ x_bx_b)-\frac{1}{2}(N^2-1)+f^{\dagger}_b f_b,
\]
\[
Q^{\dagger}=f^{\dagger}_b\pi_b+i f^{\dagger}_b x_b=\sqrt{2}i f^{\dagger}_b{a}_b, \ \ \ \ Q=-\sqrt{2}i f_b{a^{\dagger}}_b.
\]
Since the  hamiltonian is simply the number of quanta operator the
vacuum of the system is the Fock vacuum $\mid 0 \rangle$. Let us
search for the solutions of the following form
\[
\mid \psi \rangle =P_k \sum_{n=0}^{\infty} c_i (X^2)^n\mid 0 \rangle,
\]
where $P_k$s are polynomials introduced in section 3 and $c_n$s are
some coefficients.
Using the properties (7) of these polynomials and the fact that $\pi_a\mid 0 \rangle=i x_a \mid 0 \rangle$ we
find
\[
HP_k(X^2)^n\mid 0 \rangle=-n(2n+1+N^2-4+2k)P_k(X^2)^{n-1}\mid 0 \rangle+(2n+k)P_k(X^2)^n\mid 0 \rangle,
\]
hence we obtain the recurrence
\[
c_{n+1}=\frac{2n+k-E}{(n+1)(2(n+1)+1+N^2-4+2k)}c_n.
\]
where $E$ is the energy. In the standard fashion we require that $c_{i>n}=0$ in order to make
the state $\mid \psi \rangle$ properly normalized. This brings us to
the condition $E=2n+k$ which is the eigenvalue of $H$ corresponding
to the eigenvector

\[
\mid \psi \rangle =P_k H_{n}\mid 0 \rangle, \ \ \ \ H_{n}=  \sum_{i=0}^{n} c_i (X^2)^i.
\]
The leading term in $N$ of $\left\| (X^2)^i\mid 0 \rangle
\right\|^2$ is proportional to $N^{4i}$.  \footnote{The best way to
see this is in coordinate representation. We have
\[
\left\| (X^2)^i\mid 0 \rangle \right\|^2=\int dX (X^2)^{2i}
e^{-(X^2)}= \sqrt{\pi}^{2i}\left( a^{\frac{1-N^2}{2}} \right)^{(2i)}
\xrightarrow[N \longrightarrow \infty]{} \sqrt{\pi}^{2i} N^{4i}
\]
  }
On the other hand $c_i^2$s are proportional to $1/N^{4i}$ therefore
all the terms of $H_{n}$ survive the $N \to \infty$ limit in the sense of their norms hence
large $N$ solutions are given by linear combinations of $(X^2)^n(X^k)\mid 0 \rangle$. Rewriting this
conclusion in terms of creation operators acting on Fock vacuum we
obtain ( see Appendix D )
\begin{equation}
\mid k \rangle =({a^{\dagger}}^k)\mid 0 \rangle.
\end{equation}

This is exactly the basis introduced by Veneziano and Wosiek.
However it would be to hasty to conclude that such basis should be
used in the large $N$ limit whenever bound states occur. This is
because the state (26) is only one example of many solutions that appear
in supersymmetric harmonic oscillator. Since the  hamiltonian is the
number of quanta operator, the following states

\begin{equation}
\mid i_2,i_3,\ldots,i_N \rangle =\frac{1}{\mathcal{N}_{i_2,i_3,\ldots,i_N}}
({a^{\dagger}}^2)^{i_2}({a^{\dagger}}^3)^{i_3}\ldots
({a^{\dagger}}^N)^{i_N}\mid 0 \rangle,
\end{equation}
with the normalization factor $\mathcal{N}_{i_2,i_3,\ldots,i_N}$ are
the eigenstates of $H$  with the eigenvalue
$2i_2+3i_3+\ldots+Ni_N$ and (26) corresponds
to just one of them. Moreover one can prove (see Appendix D) that in
the large $N$ limit these states become orthogonal and that

\begin{equation}
\mathcal{N}_{i_2,i_3,\ldots,i_N}^2=2^{i_2}i_2! 3^{i_3}i_3!  \ldots
N^{i_N}i_N! N^{2i_2+3i_3+\ldots+Ni_N}, \ \ \ \ N \to \infty.
\end{equation}
Therefore in the sector with fixed number of quanta $n_B$ the norms
of all the states have the same $N$ dependence namely $N^{n_B}$.
\footnote{From Eqn. (28) it may seem at first sight that the longest
traces admit the additional factor of $N^{i_N}$. This is however an
illusion. If the state has $n_B$ number of quanta then $i_k$s are
fixed and do not depend on $N$, moreover starting from $k>n_B$ we have
$i_k=0$  } This result is a little bit surprising since at first
sight it seemed that the norms of the single trace states grow
faster with $N$ then the norms of any other states.  Since this is not
true, there has to be some other criterium that distinguishes their
role. However this discussion is beyond the scope of this paper.

\section{Summary}
In this paper we attempted to understand the behavior of quantum
mechanics based on $SU(N)$ group when $N$ is large but finite. The
discussion is far from complete since the models analyzed here are
the easiest ones. We have shown that when the hamiltonian  is free
$H=\frac{1}{2}\pi_a \pi_a$ then the single trace states
occur as the large $N$ limit of certain solutions. However
there are also bilinear operators $(a^{\dagger}a^{\dagger})^k$ which  should be taken into
consideration as well. Their emergence is due to the structure of
the vacuum state in Fock space so it is not clear weather bilinear
operators should be included when we are discussing other
hamiltonians with more complicated potentials. Let us also note that
in the case of supersymmetric harmonic oscillator non of the states
from the Fock space are favored. However there is no t'Hooft
coupling in this case so it is not a good example to study the $N
\to \infty$ limit.

 Another interesting issue is the
supermultiplet structure which we did not discuss here at all and
which certainly simplifies in large $N$ limit.

We also hope that the algebraic approach presented here will be
useful while analyzing SYMQM in D=4 and D=10 dimensions. In fact
this was our main motivation.

\section{Acknowledgments}
I thank G. Veneziano and J. Wosiek for the discussions. This work was supported by the
the grant of Polish Ministry of Science and Education P03B 024 27
( 2004 - 2007 ).

\section{Appendix A}

In this appendix we prove (5) and (6). First let us note that  if
$W$ is an arbitrary function of traces $(X^k)$ then $[\pi_a,
W]=-i\partial_a W \pi_a$ hence

\begin{equation}
[\pi^2, W]=-2i\partial_a(W)\pi_a -\partial_a \partial_a W.
\end{equation}
The proof of (5) is now  straightforward. The proof of (6) requires the identity
$[T_a]_{ij}[T_a]_{kl}=\delta_{il}\delta_{jk}-\frac{1}{N}\delta_{ik}\delta_{jl}$
for $SU(N)$ generators. From this identity it follows that
\begin{equation}
(AT_aBT_a)=(A)(B)-\frac{1}{N}(AB),
\end{equation}
where $A,B$ are arbitrary matrices. According to (29) we have
\begin{equation}
[\pi^2,(X^k)]=-2ik(X^{k-1}\pi)-k\sum_{j=0}^{k-2}(T_aX^jT_aX^{k-2-j}).
\end{equation}
The last term in (31) is evaluated with use of (30). It is
particularly convenient to extract all the $N$ dependence i.e.
\[
\sum_{j=0}^{k-2}(T_aX^jT_aX^{k-2-j})=(2N-\frac{k-1}{N})
(X^{k-2})+\sum_{j=2}^{k-4} (X^{k-2-j})(X^j), \ \ \ \ k>2,
\]
where the sum on the right hand side should not be included when $k<6$.

\section{Appendix B}
Here we prove the existence of polynomials  $P_k$ (7) and their
properties. First we rewrite (6) in the following form

\begin{equation}
[\pi^2, P_k]=-2i\partial_a P_k \ \pi_a + \tilde{P}_{k-2},
\end{equation}
\[
 P_k=(X^k), \ \ \ \ \tilde{P}_{k-2}=k(2N-\frac{k-1}{N})
(X^{k-2})+k\sum_{j=2}^{k-4} (X^{k-2-j})(X^j).
\]
$P_k$ is of order $k$ in $x_a$ variables  and $\tilde{P}_{k-2}$ is
of order $k-2$ in $x_a$.

We now argue that it is possible to add  to $P_k$ terms in such a
way that $\tilde{P}_{k-2}=0$. The proof of this is inductive.
First we note that for any polynomial $W_{k-2}$ of order $k-2$,
according to (5) we have
\begin{equation}
[\pi^2, (X^2)W_{k-2}]=-2i\partial_a\left((X^2)W_{k-2}\right)
\pi_a-2(N^2-1)W_{k-2}-4x_a\partial_aW_{k-2}+(X^2)\tilde{W}_{k-4},
\end{equation}
where $\tilde{W}_{k-4}=\partial_a\partial_a W_{k-2}$ is of order  $k-4$. Now we note that
polynomial $\tilde{P}_{k-2}$ in (31) does not have terms of order
lower then $k-2$. This implies that
\begin{equation}
x_a\partial_a \tilde{P}_{k-2}=(k-2)\tilde{P}_{k-2},
\end{equation}
therefore taking $W_{k-2}=\tilde{P}_{k-2}$ and putting (34) into
(33) we obtain
\begin{equation}
[\pi^2,(X^2)\tilde{P}_{k-2}]=-2i\partial_a\left[(X^2)\tilde{P}_{k-2}\right]
\pi_a-2(N^2-3+2k)\tilde{P}_{k-2}+(X^2){\tilde{P}^{(1)}}_{k-4},
\end{equation}
where ${\tilde{P}^{(1)}}_{k-4}-\partial_a\partial_a \tilde{P}_{k-2}$ is of order $k-4$.  With use of
the last equation the $\tilde{P}_{k-2}$ term in (31) can now be
subtracted. We have

\begin{equation}
[\pi^2,Q^{(1)} ]=
-2i\partial_a Q^{(1)} \ \pi_a +
 \frac{1}{2(N^2-3+2k)}(X^2){\tilde{P}^{(1)}}_{k-4},
\end{equation}
where
\[
Q^{(1)}=P_k+\frac{1}{2(N^2-3+2k)}(X^2)\tilde{P}_{k-2}.
\]
The last term in (36) did not cancel  however we see that it also
obeys (34) ( i.e if instead of $\tilde{P}_{k}$ in (34)  is
take $\frac{1}{2(N^2-3+2k)}(X^2){\tilde{P}^{(1)}}_{k-4}$ then (34)
will be true ). Therefore we may apply (33) for this term and we
obtain

\[
[\pi^2, Q ]=-2i\partial_a  Q  \ \pi_a +
 \frac{1}{2(N^2-3+2k)(N^2-5+2k)}(X^2)^2{\tilde{P}^{(2)}}_{k-6},
\]
where
\[
Q=P_k+\frac{1}{2(N^2-3+2k)}(X^2)\tilde{P}_{k-2}+
\frac{1}{4(N^2-3+4k)(N^2-5+2k)}(X^2)^2{\tilde{P}^{(1)}}_{k-4},
\]
and ${\tilde{P}^{(2)}}_{k-6}=\partial_a\partial_a \tilde{P}_{k-4}$ is of order $k-6$.  This inductive
procedure gives us the family of  polynomials $P^{(i-1)}_{k-2i}$ .
Since these polynomials are made out of traces $(X^n)$ and since
our group is $SU(N)$ we have $P^{(\frac{k-1}{2})}_{1}=0$.
Therefore the procedure described here stops only on the
polynomial $P^{(\frac{k-2}{2})}_{2}=(X^2)$. The remaining term to
subtract will be proportional to $(X^2)^{k-1}$ and this can be done
using (5).

This ends the proof of the existence of polynomials $P_k$ as well
as gives the way to construct them. The property  $P_k
\xrightarrow[N \longrightarrow \infty]{}  (X^k)$ is
evident. The identity $\Delta P_k =0$ follows from (29).

\section{Appendix C}
In this appendix we prove (17). We note
that since fermion operators $f^{\dagger}_a$ anticommute we have
\[
f^{\dagger}f^{\dagger}=(f^{\dagger}f^{\dagger}T_a)T_a, \ \ \ \ f^{\dagger}=f^{\dagger}_aT_a,
\]
therefore
\[
({f^{\dagger}}^{2n+1})= (f^{\dagger}f^{\dagger}T_{a_1})\ldots
(f^{\dagger}f^{\dagger}T_{a_n})(T_{a_1}\ldots T_{a_n} F).
\]
Since operators $(f^{\dagger}f^{\dagger}T_{a_k})$ commute we may
symmetrize over indices

\[
({f^{\dagger}}^{2n+1})=\frac{1}{n!}
(f^{\dagger}f^{\dagger}T_{a_1})\ldots
(f^{\dagger}f^{\dagger}T_{a_n})(T_{(a_1}\ldots T_{a_n)} F).
\]
Generators $T_a$ are $N \times N$ matrices therefore  according to
Cayley-Hamilton theorem if $n \ge N$ then matrix $T_{(a_1}\ldots
T_{a_n)}$ can be expressed as a linear combination of products of
matrices $T_{(a_1}\ldots T_{a_k)}$, $k<n$ \footnote{ In this way
Cayley-Hamilton theorem provides a number of identities for
structure tensors which are very well known [15]. For a
compilation of identities involving invariant tensors see [16]}.
This implies that operators $({f^{\dagger}}^{2n+1})$, $n \ge N$
 can be expressed as a linear combination of products of
operators $({f^{\dagger}}^{2n+1})$, $n<N$. In fact we have
proven something even more general i.e. that for fermionic
matrices $f^{\dagger}$ there is a Cayley-Hamilton theorem
expressing ${f^{\dagger}}^{2n+1}$, $n \ge N$ in terms of
${f^{\dagger}}^{2n+1}$, $n < N$.

\section{Appendix D}
Here we prove Eqns. (23), (26) and (28). In order to prove (23) we note that
\[
(X^k)({a^{\dagger}}^2)^n \mid 0 \rangle= \sum_{i_k, j_k}
(a^{i_1}{a^{\dagger}}^{j_1}  \ldots
a^{i_n}{a^{\dagger}}^{j_n})({a^{\dagger}}^2)^n\mid 0 \rangle,
\]
however if $A$ is an arbitrary  operator valued matrix and
$k,n=0,1,2, \ldots$ then
\[
(Aa{a^{\dagger}}^k)({a^{\dagger}}^2)^n\mid 0 \rangle=
\sum_{j=0}^{k-1}(AT_i{a^{\dagger}}^jT_i{a^{\dagger}}^{k-2-j})({a^{\dagger}}^2)^n
\mid 0 \rangle +2n(A{a^{\dagger}}^k)({a^{\dagger}}^2)^{n-1}\mid 0
\rangle \xrightarrow[N \longrightarrow \infty]{}
\]
\[
\xrightarrow[N \longrightarrow \infty]{} N(A{a^{\dagger}}^{k-1})({a^{\dagger}}^2)^n\mid 0\rangle,
\]
therefore in the large $N$ limit the  leading term of
$(Aa{a^{\dagger}}^k)({a^{\dagger}}^2)^n\mid 0 \rangle$ is
$$(A{a^{\dagger}}^{k-1})({a^{\dagger}}^2)^n\mid 0 \rangle.$$ Since
A is  arbitrary the leading term of $(X^k)({a^{\dagger}}^2)^n\mid
0 \rangle$ is $({a^{\dagger}}^k)({a^{\dagger}}^2)^n\mid 0
\rangle$.

Eqn. (26) is surprising since there are no bilinear operators
$(a^{\dagger}a^{\dagger})$ but indeed it is true. Let us start with
the simplest, illustrative case
\[
(X^2)\mid 0 \rangle=\frac{1}{2}\left(({a^{\dagger}}^2)+N^2-1 \right)\mid 0 \rangle
\]
therefore the norm is
\[
\langle 0 \mid(X^2)^2\mid 0 \rangle \xrightarrow[N \longrightarrow \infty]{} \frac{1}{4}N^4.
\]
Note that the $N^4$ dependence does not come from the norm of the
state $({a^{\dagger}}^2)\mid 0 \rangle$ but as a square on $N^2$
i.e. from the vacuum state. This picture continues. Let us take
\[
(X^2)^k\mid 0 \rangle=\frac{1}{2^k}\left(({a^{\dagger}}^2)+({a}^2)+(a^{\dagger}a)+N^2-1 \right)\mid 0 \rangle
\]
and expand it in terms of powers of
$({a^{\dagger}}^2)$,$({a}^2)$,$(a^{\dagger}a)$.  There is one term
which does not have any  of those operators namely
$\frac{1}{2^k}(N^2-1)^k$. We now argue that it is leading in $N$.

The operators $(a^{\dagger}a)$ count only the number of  quanta so
after commuting them in front of  $\mid 0 \rangle$ we do not get any
additional factors dependent on $N$. Therefore we are left with
various powers of operators $({a^{\dagger}}^2)$ and $({a}^2)$. In
order to commute $({a}^2)$ in front of $\mid 0 \rangle$ we use the
relation
\[
[(a^2),({a^{\dagger}}^2)^n]=4n({a^{\dagger}}^2)^{n-1}(a^{\dagger}a)+2n(2n-2+N^2-1)({a^{\dagger}}^2)^{n-1}.
\]
Therefore to each such commutation we assign the $N^2$ factor. There
are no more then $k-1$ such commutations so at most we get the
$N^{2k-2}$ factor. Therefore we can write
\[
(X^2)^k\mid 0 \rangle \xrightarrow[N \longrightarrow \infty]{}\frac{1}{2^k}(N^2-1)^k\mid 0 \rangle.
\]
In other words, among all the vectors from the Fock space  the
greatest contribution to the state $(X^2)^k\mid 0 \rangle$ comes
from the Fock vacuum. Therefore the greatest contribution to the
state $(X^n)(X^2)^k\mid 0 \rangle$ comes from  $(X^n)\mid 0 \rangle$
the leading term of which is $({a^{\dagger}}^n)\mid 0 \rangle$.

To prove (28) we start with the observation that if

\[
\mid s \rangle=t_{i_1 \ldots i_k}{a^{\dagger}}_{i_1}\ldots
{a^{\dagger}}_{i_k}\mid 0 \rangle, \ \ \ \ t_{i_1 \ldots i_k} \in
\textbf{C},
\]
then
\begin{equation}
\langle s \mid s \rangle = \sum_{\sigma \in S_k}
t^{*}_{\sigma(i_1)  \ldots \sigma(i_k)} t_{i_1 \ldots
i_k}\delta_{\sigma(i_1)i_1}\ldots \delta_{\sigma(i_k)i_k} =
t^{*}_{(i_1 \ldots i_k)} t_{i_1 \ldots i_k}.
\end{equation}
In our case we have
\[
\mid s \rangle=({a^{\dagger}}^2)^{i_2}({a^{\dagger}}^3)^{i_3}\ldots
({a^{\dagger}}^N)^{i_N}\mid 0 \rangle,
\]
therefore $t_{j_1 \ldots j_k}$ is the product of traces $(T_{j_1} T_{j_2}\ldots T_{j_k})$
\[
t_{j_1 \ldots j_k}=(T_{j_1} T_{j_2})\ldots
(T_{j_{2i_2+1}}T_{j_{2i_2+2}}T_{j_{2i_2+3}})\ldots
(T_{j_{2i_2+3i_3+1}}T_{j_{2i_2+3i_3+2}}T_{j_{2i_2+3i_3+3}}T_{j_{2i_2+3i_3+4}})\ldots \ \ \ .
\]
Among all the permutations in the sum (37) there is exactly one  (
the identity $\sigma=id$) which, due to Kronecker deltas in (37),
contracts $i_1$ with $\sigma(i_1)$,  $i_2$ with $\sigma(i_2)$ etc.
This term, as we shall see, gives the contribution to the leading
term in $N$. There are other terms with the same $N$ dependence but we
will include them later on.

 Since
$T_i's$ are hermitian we have $(i_1\ldots i_k)^*=(i_k\ldots i_1)$
therefore we write

\begin{equation}
\langle s \mid s \rangle =   t_{i_k \ldots i_1}  t_{i_1 \ldots
i_k}  + (\hbox{the rest of permutations}).
\end{equation}
If we introduce the following notation
\[
s_k:=(j_k \ldots j_1)(j_1 \ldots j_k),
\]
then it follows that
\begin{equation}
t_{i_k \ldots i_1} t_{i_1 \ldots i_k}=s_2^{i_2}\ldots s_k^{i_k}.
\end{equation}
The computation of $s_k$ is straightforward. With use of the to
the identities

\begin{equation}
(A T_j)(B T_j)=(AB)-\frac{1}{N}(A)(B),  \ \ \ \ T_{j}T_{j}=(N-\frac{1}{N})\bold{1},
\end{equation}
we have
\[
s_k=(T_{j_k}\ldots T_{j_1})(T_{j_1}  \ldots
T_{j_k})=(T_{j_k}\ldots T_{j_2} T_{j_2} \ldots
T_{j_k})-\frac{1}{N}(T_{j_k}\ldots T_{j_2})(T_{j_2} \ldots
T_{j_k}).
\]
The first term on the right hand side of the above equation is
\[
(T_{j_k}\ldots T_{j_2} T_{j_2} \ldots T_{j_k})
=(N-\frac{1}{N})(T_{j_k}\ldots T_{j_3} T_{j_3} \ldots
T_{j_k})=\ldots
\]
\[
\ldots=(N-\frac{1}{N})^{k-2}(T_{j_k}
T_{j_k})=(N-\frac{1}{N})^{k-1}N.
\]
The second term is simply $-\frac{1}{N}s_{k-1}$ therefore
\[
s_k=(N-\frac{1}{N})^{k-1}N-\frac{1}{N}s_{k-1},
\]
hence in the large $N$ limit we have $s_k= N^{k}$. From (38,39) it follows now that

\begin{equation}
\langle s \mid s \rangle =
N^{2i_2+3i_3+\ldots+Ni_N}  + (\hbox{the rest of permutations}).
\end{equation}
Now  we will extract from the last term in (41) "$( \hbox{the rest
of permutations} )$" all the other leading terms. Due to the
cyclicity of traces for each trace $(T_{j_1} \ldots T_{j_k})$ there
are $k$ different permutations that give the same term as for the
identity permutation. Therefore if there are $i_2$ traces of the
length 2, $i_3$ traces of the length 3, etc. then there are
$2^{i_2}3^{i_3} \ldots N^{i_N}$ additional permutations that give
the leading terms. Moreover, among the group of the traces of the
length $k$ we can permute those traces in $i_k!$ ways obtaining
additional permutations that give the leading terms. Therefore there
are in fact $2^{i_2}i_2!3^{i_3}i_3! \ldots N^{i_N}i_N!$ leading
terms. Each of those terms is equal to the term with $\sigma=id$ so
that they are equal to $N^{2i_2+3i_3+\ldots+Ni_N}$.  Therefore the
leading term of (41) is now
\begin{equation}
\langle s \mid s \rangle =
2^{i_2}i_2!3^{i_3}i_3! \ldots N^{i_N}i_N!N^{2i_2+3i_3+\ldots+Ni_N}  + (\hbox{the rest of permutations'}).
\end{equation}
where the ' means  all the permutations but the ones we have just extracted.

It is now sufficient to prove that the last term in (42)
"$(\hbox{the rest of permutations'})$"   grows slower with $N$ then
$N^{2i_2+3i_3+\ldots+Ni_N}$. To see this we note that each
contraction of one generator $T_a$ with another one may give at
best the factor of N. This happens only if the contracted
generators are next to each other i.e.
\begin{equation}
(AT_aT_aB)=(N-\frac{1}{N})(AB) \xrightarrow[N \longrightarrow \infty]{} N(AB),
\end{equation}
where A and B are arbitrary matrices. If the generators are separated by another matrix $C$ then
\begin{equation}
(AT_aCT_aB)=(AB)(C)-\frac{1}{N}(ACB).
\end{equation}
We see that if $C=1$ then we recover the previous result but when $C
\ne 1$ then we do not gain the $N$ factor. Therefore among all the
possible permutation in (37) the $\sigma=id$ and the equivalent
ones, that we have extracted, are leading in $N$ because in
evaluating the term with $\sigma=id$ we repeatedly used (43) and
never used (44). There are no other terms leading in $N$ since there
is at least one contraction of the form (44) with $C \ne 1$ in each
term of "$(\hbox{the rest of permutations'})$". Therefore we finally
get
\[
\langle s \mid s \rangle=2^{i_2}i_2!3^{i_3}i_3! \ldots N^{i_N}i_N! N^{2i_2+3i_3+\ldots+Ni_N} \ \ \ \ N \to \infty.
\]

Finally, prove that the states (27) become orthogonal  in $N \to
\infty$ limit. Let us consider two arbitrary states
\[
\mid i_2,i_3,\ldots,i_N \rangle=t_{i_1 \ldots
i_{n_B}}{a^{\dagger}}_{i_1}\ldots {a^{\dagger}}_{i_{n_B}}\mid 0
\rangle,
\]
\begin{equation}
\mid j_2,j_3,\ldots,j_N \rangle= t'_{i_1 \ldots
i_{n_B}}{a^{\dagger}}_{i_1}\ldots {a^{\dagger}}_{i_{n_B}}\mid 0 \rangle,
\end{equation}
with the number of quanta $n_B$.  According to Cauchy-Schwarz
inequality we have
\[
\langle i_2,i_3,\ldots,i_N \mid j_2,j_3,\ldots,j_N \rangle \propto N^{l}, \ \ \ \ l \le n_B.
\]

We have already shown that the case  $l=n_B$ corresponds to the
norms of states with $n_B$ number of quanta. The question remains
weather there are scalar products, of two distinct states, that have
the same $N$ dependence. Fortunately the answer is no. To see this
we write the scalar product as
\begin{equation}
\langle i_2,i_3,\ldots,i_N \mid j_2,j_3,\ldots,j_N \rangle=t^{*}_{(i_1 \ldots i_{n_B})} t'_{i_1 \ldots i_{n_B}}.
\end{equation}
Since $t_{(i_1 \ldots i_{n_B})}$  and $t'_{i_1 \ldots i_{n_B}}$  are
now different each term in the sum (46) will have at least one
contraction of the form \footnote{ The best way to read (47) is to
start with the $\times$ in the middle and read the left and right
hand side which are contracted. The integers p/q are the numbers of
traces on the left/right hand side of $\times$. }

\[
s_k^{p \ q }=(T_{i_{k}} \ldots
T_{i_{m_{p-1}+1}})\ldots(T_{i_{m_2}}  \ldots
T_{i_{m_1+1}})(T_{i_{m_1}} \ldots T_{i_1}) \times
\]
\begin{equation}
(T_{i_1} \ldots T_{i_{n_1}})
(T_{i_{n_1+1}} \ldots T_{i_{n_2}}) \ldots (T_{i_{n_{q-1}+1}}
\ldots T_{i_k}).
\end{equation}

Naturally, such contractions differ from $s_k=s_k^{1 \ 1}$'s therefore
we cannot express the leading term of (46) in terms of $s_k$'s
like in (39). Moreover, using the identities (40) and
performing the same manipulations it follows that
\[
s_k^{p \ q}=N^{k-(p-1)-(q-1)}, \ \ \ \ N \to \infty,
\]
therefore for $p=q=1$ we recover the previous  result
$s_k^{1 \ 1}=s_k=N^k$ but when $p>1$ or $q>1$ then the N dependence will be
slower. Since the leading term of (45) always contains at least
one $s_k^{p \ q}$ with $p>1$ or $q>1$ it follows that
\[
\langle i_2,i_3,\ldots,i_N \mid j_2,j_3,\ldots,j_N \rangle \propto N^{l}, \ \ \ \ l < k.
\]

The states  (45) become orthogonal, in the large $N$ limit, in the following sense. Since

\begin{equation}
\langle i_2,i_3,\ldots,i_N \mid j_2,j_3,\ldots,j_N   \rangle
\propto N^k, \ \ \ \ k<n_B, \ \ \ \ N \to \infty,
\end{equation}
therefore if we define
\[
\mid i_2,i_3,\ldots,i_N \rangle'=\frac{1}{\sqrt{2^{i_2}i_2!
3^{i_3}i_3! \ldots N^{i_N}i_N!  N^{2i_2+3i_3+\ldots+Ni_N}}}\mid
i_2,i_3,\ldots,i_N \rangle,
\]
then we have
\[
'\langle i_2,i_3,\ldots,i_N \mid j_2,j_3, \ldots,j_N \rangle'=\delta_{i_2 j_2} \delta_{i_3 j_3}\ldots
\delta_{i_N j_N} \ \ \ \ N \to \infty.
\]

\end{document}